\documentclass[%
reprint,
superscriptaddress,
 amsmath,amssymb,
 aps,
floatfix,
]{revtex4-1}

\usepackage[dvipsnames,svgnames,x11names,hyperref]{xcolor}
\usepackage{siunitx}

\usepackage{graphicx}
\usepackage{dcolumn}
\usepackage{bm}
\usepackage{hyperref}
\usepackage[mathlines]{lineno}
\usepackage{mathrsfs}
\usepackage[normalem]{ulem}

\usepackage[margin=0.92in]{geometry}

\usepackage{braket}

\newcommand{\cmmb}[1]{\textcolor{blue}{#1}}
\usepackage[normalem]{ulem}
\hypersetup{colorlinks=true, 
	linkcolor={blue!75!black!80!yellow},
	citecolor={blue!75!black!80!yellow}, 
	urlcolor=black 
	}
\makeatletter \renewcommand\@make@capt@title[2]{%
\@ifx@empty\float@link{\@firstofone}{\expandafter\href\expandafter{\float@link}}%
\sffamily{\textbf{#1}}\@caption@fignum@sep#2 }

\begin{document}
\preprint{APS/123-QED}

\title{Weak-to-Strong Light-Matter Coupling and \\
Dissipative Dynamics from First Principles}

\author{Derek S. Wang}
\email{derekwang@g.harvard.edu}
\affiliation{Harvard John A. Paulson School of Engineering and Applied Sciences, Harvard University, Cambridge, MA 02138, USA}
\author{Tom\'{a}\v{s} Neuman}
\affiliation{Harvard John A. Paulson School of Engineering and Applied Sciences, Harvard University, Cambridge, MA 02138, USA}
\author{Johannes Flick}
\affiliation{Center for Computational Quantum Physics, Flatiron Institute, New York, NY 10010, USA}
\author{Prineha Narang}
\email{prineha@seas.harvard.edu}
\affiliation{Harvard John A. Paulson School of Engineering and Applied Sciences, Harvard University, Cambridge, MA 02138, USA}

\begin{abstract}
Cavity-mediated light-matter coupling can dramatically alter opto-electronic and physico-chemical properties of a molecule. \textit{Ab initio} theoretical predictions of these systems need to combine non-perturbative, many-body electronic structure theory-based methods with cavity quantum electrodynamics and theories of open quantum systems. Here we generalize quantum-electrodynamical density functional theory to account for dissipative dynamics and describe coupled cavity-molecule interactions in the weak-to-strong-coupling regimes. Specifically, to establish this generalized technique, we study excited-state dynamics and spectral responses of benzene and toluene under weak-to-strong light-matter coupling. By tuning the coupling we achieve cavity-mediated energy transfer between electronic excited states. This generalized \emph{ab initio}  quantum-electrodynamical density functional theory treatment can be naturally extended to describe cavity-mediated interactions in arbitrary electromagnetic environments, accessing correlated light-matter observables and thereby closing the gap between electronic structure theory and quantum optics.

\end{abstract}
\date{\today}

\maketitle

\section{Introduction}

Recent experiments have explored regimes of light-matter interaction of molecular systems in nanophotonic cavities, where the interaction of a photon with the excitations of a single molecule can be substantially enhanced. When the loss in the cavity, characterized by a loss rate $\kappa$, dominates over the light-matter coupling rate $g$ ($g<\kappa$) the system is in the weak-coupling regime. In this regime the cavity mode enhances light-matter interactions, which can result in decay enhancement of molecular excited states via the Purcell effect~\cite{purcell1946spontaneous}. This regime has been leveraged to improve spectroscopic detection of molecular species~\cite{wrigge2008efficient, kinkhabwala2009large, aroca2013plenhspec, pelton2015modified, itoh2017reviewempl} without qualitatively modifying the electronic character of the molecular excitations.

A qualitative change in the physico-chemical properties of matter emerges when 
$g>\kappa$, which leads to the hybridization of excitations of matter with cavity photons to form polaritonic states~\cite{cwik2016excitonic, ebbesen2016strongcoupling, flick2017, OwrutzkyReview, rivera2018, flick2018b, flick2018strong, flickexcited}. This coherent non-perturbative regime is denoted as strong light-matter coupling~\footnote{For more detailed discussions of the definition of light-matter coupling regimes, see Ref.~\cite{torma2014strong, melnikau2016rabi, zengin2016conditionssc, autore2018boron, flick2018strong}}, resulting in modifications of molecular chemical reactivity via electronic or vibrational strong coupling~\cite{galego2015, galego2016, anoop2016vibreactivity, herrera2016chemistry, flick2017ab, galego2019, groenhof2019relaxation, ebbesenTilting}, electrical~\cite{ebbesenConductivity} and excitonic~\cite{feistExcitonConductivity} conductivity, optical properties~\cite{lidzey1999polemission, delpino2015quantum, george2015ultra, herrera2018review, zeb2018exactvibdressed, herrera2017dark, Owrutzsky2DIR}, energy transfer~\cite{zhong2017entangled, coles2014polariton, juraschek2019cavity, DuEnergyTransfer}, polariton lasing~\cite{kenacohenLasingExp,kenacohenLasingTheory}, or superconductivity~\cite{Laussy2010,Schlawin2019,Curtis2019}.

In the intermediate regime, when $g\sim \kappa$,  the distinction between weak and strong light-matter coupling vanishes. This regime has been shown in optical cavities characterised by small quality factor $Q$, where strong confinement of light in plasmonic~\cite{neuman2018coupling} or conventional~\cite{hagenmuller2019adiabatic} cavities can lead to efficient coupling of light with excitons in many~\cite{zengin2013approaching, gulis2015realizing, wersall2017strongcplpl} or a few organic molecules~\cite{chikkaraddy2016single}. Here the polaritonic states are no longer well defined long-lived states~\cite{silva2019polaritonic} nor can the light-matter interaction be treated as a weak perturbation to molecular excitations. This regime has been suggested to influence incoherent dissipative rates governing steady-state populations of molecular electronic excited states and can, for example, protect organic molecules from photo-oxidation processes~\cite{Munkhbat2018suppression}.

Theoretical descriptions of coupling of light with molecular excitations often rely on models based on cavity quantum electrodynamics (cQED) and theories of open quantum systems~\cite{Breuer2005, blum2013density}. 
Such models require the parametrization of input Hamiltonians and loss terms describing the dynamics of both the excitations and the cavity field. 
\textit{Ab initio} predictions of the effects of light-matter coupling on the molecular properties therefore require non-perturbative methods combining the power of cQED, theories of open-quantum systems, and many-body electronic structure theory~\cite{rivera2018}. 

To bridge this critical gap in the field, we build on the formalism of quantum-electrodynamical density functional theory (QEDFT)~\cite{tokatly2013,ruggenthaler2014,Dimitrov_2017,flick2018b, Ruggenthaler2018, flick2019lmrnqe, flick2018strong, flickexcited}, described in Section \ref{ssec:hamiltonian}, to incorporate cavity losses. We apply our generalized formalism to study \textit{ab initio} vacuum light-matter interaction in both the weak- and strong-coupling regimes. Concretely, we study an isolated electronic transition in a benzene molecule in Section \ref{ssec:twolevel} and a toluene molecule in Section \ref{ssec:manylevel}, where many electronic transitions lead to an effective many-level electronic system. We show that light-matter interactions can induce interactions among the electronic states of the molecules that can result in energy transfer between excited states.

\section{Formalism and Methodology} \label{ssec:hamiltonian}
The Hamiltonian $H$ for a non-relativistic system of $M$ electronic excited states interacting with the quantized light field of $N$ photon modes, in the absence of an external classical current and under the dipole approximation in the length gauge, is given by~ \cite{tokatly2013, flick2017, flickexcited}

\begin{align} \label{eq:hamiltonian}
H = H_e + \sum_{k=1}^{N} \frac{1}{2}\left[ p_k^2 + \omega_k^2 \left(q_k-\frac{\bm{\lambda}_k}{\omega_k}\cdot \bm{R}\right)^2\right],
\end{align}
where $H_e$ is the electronic Hamiltonian, and the $k$th quantized photon mode is given by the operators for the photon conjugate momentum $p_k={\rm i} \sqrt{\frac{\hbar\omega_k}{2}}(a_k-a_k^\dagger)$ and the photon displacement coordinate $q_k=\sqrt{\frac{\hbar}{2\omega_k}}(a_k+a_k^\dagger)$, where $\hbar$ is the reduced Planck constant. Both photon operators are given in terms of the photon annihilation $a_k$ and creation $a_k^\dagger$ operators, and $\omega_k$ is the frequency of mode $k$. The electronic system couples to the photon modes through the total position operator $\bm{R}=\sum_{i=1}^{M}\bm{r}_i$ of the electronic system and $q_k$ of the photonic system. This coupling is scaled by the cavity strength $\bm{\lambda}_k$, which is given by

\begin{align}
    \bm{\lambda}_k=\sqrt{\frac{2}{\hbar \omega_k}}\bm{E}_ke,
\end{align}  
where $\bm{E}_k$ is the amplitude of the electric field at the center of the electronic charge density
and $e$ is the elementary charge. In a lossless cavity, $\bm{\lambda}_k$ represents a set of discrete modes whose excitation energies are well separated and thus usually only one or a few modes need to be considered in the description of light-matter coupling.

To calculate $\bm{\lambda}_k$ for a lossy cavity, we replace the cavity strength of the single mode of a lossless 3D cavity $\bm{\lambda}_{\rm c}$ at energy $\hbar\omega_{\rm c}$ with many photon modes by broadening with a Lorentzian profile $L(\Delta\omega,\kappa,\omega_{k{\rm c}})$:
\begin{align} \label{eq:lorentz}
    |\bm{\lambda}_k|^2=|\bm{\lambda}_{\rm c}|^2L(\Delta\omega,\kappa,\omega_{k{\rm c}}),
\end{align}
where 
\begin{align}
    L(\Delta\omega,\kappa,\omega_{k{\rm c}})=\Delta\omega\frac{1}{2\pi}\frac{\kappa}{(\omega_{k{\rm c}})^2+(\kappa/2)^2}.
\end{align}

The Lorentzian filter distributes the total cavity strength $\bm{\lambda}_{\rm c}$ over a range of frequency values $\omega_k$ with a Lorentzian lineshape around the central mode frequency ($\omega_{k{\rm c}}=\omega_k-\omega_{\rm c}$), where the degree of broadening is controlled by the loss rate $\kappa$. Here, the modes are uniformly spaced, that is the frequency spacing  $\Delta\omega=\omega_{k+1}-\omega_k$ for all $k$ is constant. In general, transforming to or from another energy spacing $\Delta\omega(\omega_k)$ may offer computational benefits for an arbitrary electromagnetic environment; the detailed mathematical formalism is given in Appendix \ref{app:transform}.
Equation\,\eqref{eq:lorentz} also implies that the cavity strength obeys the sum rule:


\begin{align}
    \int {\rm d}\omega D(\omega) |\bm{\lambda}_k|^2 = |\bm{\lambda}_c|^2,
\end{align}
where $\Delta\omega(\omega)\rightarrow D(\omega){\rm d}\omega$ in the continuum limit, $D(\omega)$ is the density of photon modes, and ${\rm d}\omega \rightarrow 0$.

In the following, we also define the coupling rate $g_{i,k}$ that relates to the cavity strength $\bm{\lambda}_k$ and is a measure of the coupling strength between the cavity mode $k$ and a particular electronic excitation $i$ connecting the electronic ground state $|{\rm g}\rangle$ with an electronic excited state $|{\rm e}_i\rangle$ of the molecule: 
\begin{align} \label{eq:g}
    g_{i,k}=-\frac{e}{\hbar}\bm{E}_k\cdot \langle {\rm g}|\bm{R}|{\rm e}_i\rangle=-\sqrt{\frac{\omega_k}{2\hbar}}\frac{\bm{\lambda}_k}{e}\cdot \boldsymbol{d}_i.
\end{align}
Here $\bm{d}_i=e\langle {\rm g}|\bm{R}|{\rm e}_i\rangle$ are transition dipole moments which can be, for example, obtained together with their corresponding excitation energies $\hbar\omega_i$ from a standard linear-response time-dependent density-functional theory (TDDFT)~\cite{casida1995time}. These quantities can then be used as input parameters for cQED models describing light-matter interactions. This effective interaction rate naturally emerges in parametric cQED models, such as the Jaynes-Cummings  model~\cite{shore1993}. As described in Appendix \ref{app:fano}, we adapt these approaches by including the Lorentzian profile of the cavity strength to formulate a Fano-type cQED model. In this cQED model we describe the light-matter coupling via a many-mode Jaynes-Cummings term (adopting the rotating-wave approximation), and neglect the term $\propto \bm{R}^2$ appearing in the full light-matter coupling Hamiltonian given in Eq.\,\eqref{eq:hamiltonian}. We note that several studies \cite{flick2017, schaferR2, rokajR2, DebenardisR2, LiR2} have analyzed the effect of this term, which becomes in particular relevant in the ultrastrong coupling limit. In this paper, we compare our full QEDFT calculations with $\bm{R}^2$ contributions included to the cQED model without $\bm{R}^2$ contributions and comment on their influence.

Within QEDFT the electronic and photonic excitations, in the presence of nuclei, and their interactions are treated on the same quantized footing~\cite{tokatly2013,ruggenthaler2014, Ruggenthaler2018, flick2019lmrnqe}. In particular, in this \emph{Article} we adopt a linear-response time-dependent QEDFT that generalizes the Casida equation of TDDFT, first introduced in Refs.\,\cite{flick2019lmrnqe, flickexcited}. 

\section{Results and Predictions}
In Section \ref{ssec:twolevel}, we first demonstrate the present method on a model system by coupling an isolated electronic excitation of benzene to a cavity with varying cavity strength and losses. We note the qualitative similarities between the results calculated from first principles and those expected from the well-known, parametric Jaynes-Cummings model~\cite{shore1993}. In Section \ref{ssec:manylevel}, we study a more complex system that consists of many electronic states of toluene coupled to a cavity with varying cavity strength and loss, respectively. Importantly, we observe interactions between electronic excited states mediated by the cavity that can result in population transfer between excited states. 
\subsection{An isolated electronic excitation of benzene} \label{ssec:twolevel}
\begin{figure}[h!]
\centering
\includegraphics[width=1.0\linewidth]{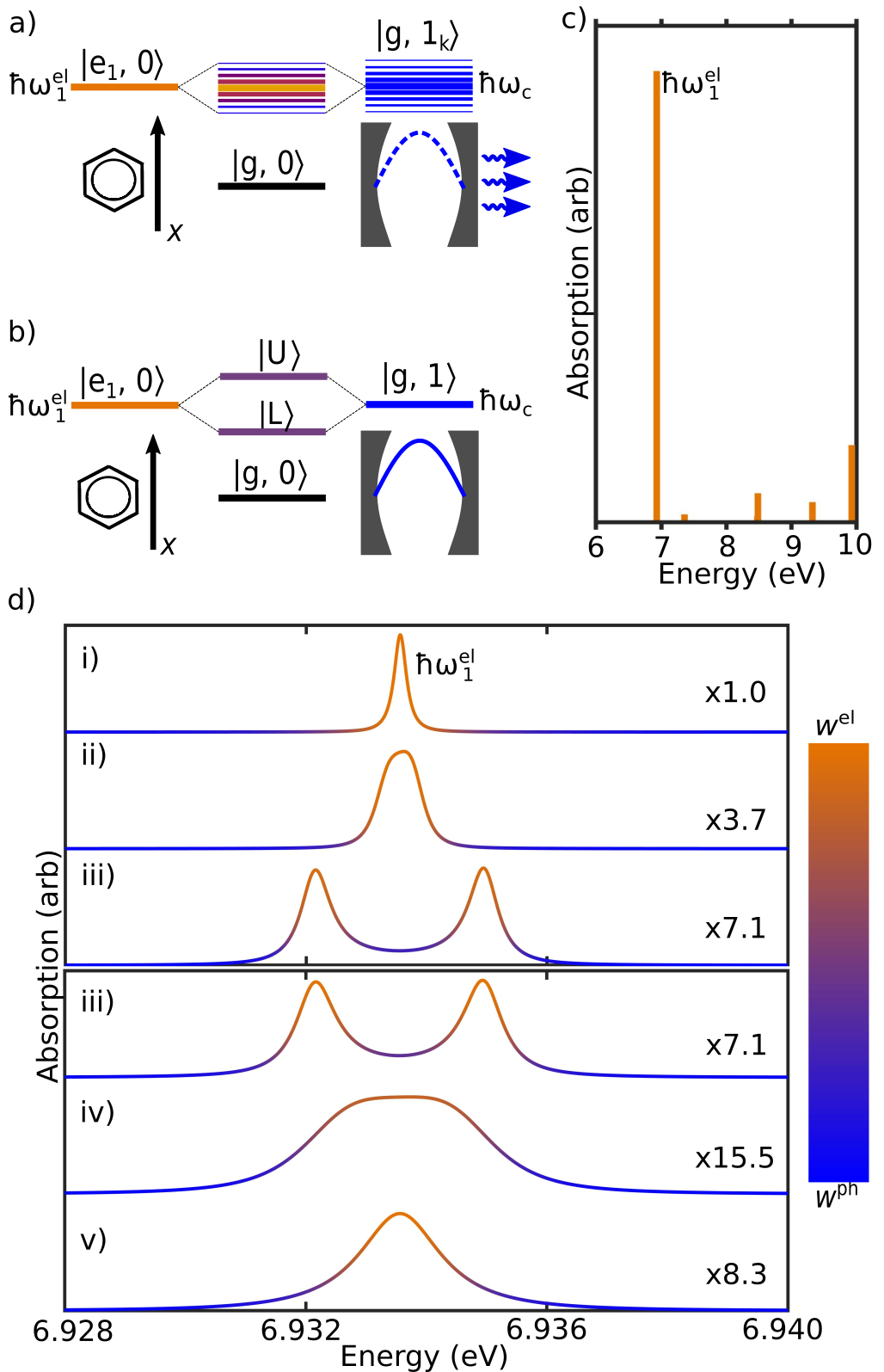}
\caption{The ground state $\ket{{\rm g}, 0}$ and first excited electronic state $\ket{{\rm e}_1, 0}$ of benzene with energy $\hbar\omega_1^\text{el}=6.93$ eV and transition dipole moment $d_{1,x}=0.96$ e\AA~comprise an effective two-level electronic system. \textbf{(a)} Coupling  $\ket{{\rm e}_1, 0}$ with lossy cavity modes $\ket{{\rm g}, 1_k}$ centered at $\hbar\omega_{\rm c}=\hbar\omega_1^\text{el}$ results in unresolved polaritonic states. \textbf{(b)} $\ket{{\rm e}_1, 0}$ couples with a resonantly tuned, low loss optical cavity $\ket{{\rm g},1}$ to form distinct upper and lower polaritonic states. \textbf{(c)} Normalized optical \textit{x}-polarized absorption spectrum of benzene in the absence of light-matter coupling. \textbf{(d)} Normalized optical \textit{x}-polarized absorption spectrum of benzene in the presence of light-matter coupling for spectral broadening $\hbar\Gamma=0.001$ eV. As the cavity strength $\lambda_{\rm c}$ increases from (i) 0.001 to (ii) 0.002 to (iii) 0.008 eV$^\frac{1}{2}$/nm with loss $\hbar\kappa=0.001$ eV in (i)-(iii), the energy splitting between the lower and upper polaritons increases; they are unable to be resolved after increasing $\hbar\kappa$ from (iii) 0.001 to (iv) 0.004 to (v) 0.008 eV with $\lambda=0.008$ eV$^{\frac{1}{2}}$/nm in (iii)-(v). The color of the curve corresponds to the relative photonic and electronic weight of the polaritonic eigenstate. To retrieve the unnormalized absorption spectra, multiply the absorption intensities by the scaling factor on the right side of the plot.}
\label{fig:1}
\end{figure}

\begin{figure}[tbhp]
\centering
\includegraphics[width=1.0\linewidth]{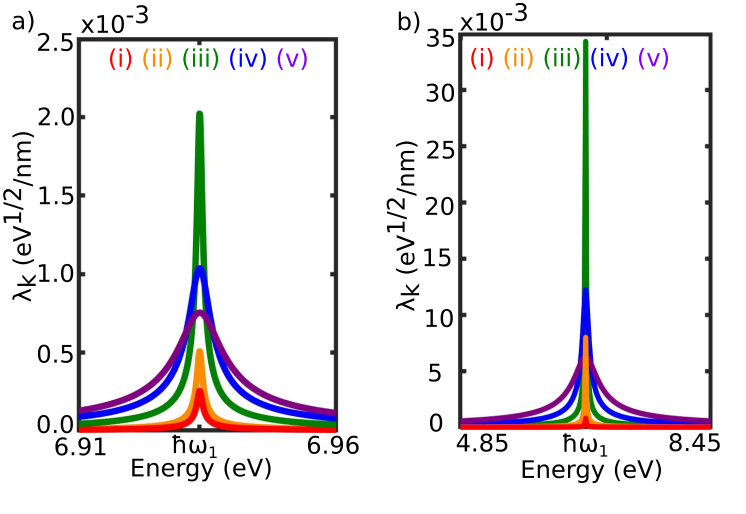}
\caption{The cavity strength $\lambda_k$ with constant frequency spacing $\Delta\omega=0.0001$ eV of the cavity modes for curves (i)-(v) for \textbf{(a)} benzene in Fig. \ref{fig:1}(d) and \textbf{(b)} toluene in Fig. \ref{fig:2}(d). For toluene in Fig. \ref{fig:3}(a)-(b), the cavity mode energy range is truncated to 6.35 eV to 6.95 eV with $\Delta\omega=0.001$ eV.}
\label{fig:cavityModes}
\end{figure}

To demonstrate our generalized QEDFT approach, we consider an example of a single benzene molecule placed in a lossy 1D optical cavity with linear polarized electric field along the $x$-axis. The orientation of the molecule is shown in Fig. \ref{fig:1}(a)-(b). In the first principles QEDFT calculation, we consider the full electronic structure with all single-particle excitations. 
Nevertheless, since the target excitation $|{\rm e}_1, 0\rangle$ with energy $\hbar\omega_1^{\rm el}=6.93$ eV and strong $x$-polarized transition dipole moment $d_{1,x}=0.96$ e\AA~tuned in resonance with the cavity mode is well separated in energy from other optically active excited states, benzene can be effectively understood as a two-level electronic system composed of the ground state $\ket{{\rm g}, 0}$ and an excited state $\ket{{\rm e}_1, 0}$ for the range of parameters we consider.
We verify this simplification by calculating an electronic $x$-polarized absorption spectrum $A_x$ of the molecule in the framework of linear-response QEDFT~\cite{flick2019lmrnqe, flickexcited}:
   \begin{align}
     A_{x}(\hbar\omega)= \mathcal{C}\sum_{l=1}^{M+N}\delta(\omega-\omega_l)\hbar\omega_l |\sum_{i=1}^{M} C_{il}^{\rm el} {d}_{i,x}|^2,
 \end{align}
where $\mathcal{C}=2m_{\rm e}/(3\hbar^2)$ (with $m_{\rm e}$ the electron mass) is a frequency-independent pre-factor, $\hbar\omega_l$ is the mode energy, $d_{i,x}$ is the $x$-component of the transition dipole moment of an electronic excitation $i$, and $C_{il}^{\rm el}$ ($C_{kl}^{\rm ph}$) is the projection of an original, unmixed electronic (photonic) state $|{\rm e}_i,0\rangle$ ($|{\rm g},1_k\rangle$) to a resulting polaritonic state $|v_l\rangle$. $M$ excited electronic states of benzene are coupled with $N$ photon modes to produce $M+N$ hybrid electron-photon states called polaritons. In the continuum limit, we broaden the delta function $\delta(\omega-\omega_l)$ with a Lorentzian:

\begin{align}
    \delta(\omega-\omega_l)\rightarrow \frac{1}{2\pi}\frac{\Gamma}{(\omega-\omega_l)^2+(\Gamma/2)^2}.
\end{align}
In Appendix \ref{app:resolution} we show the influence of the spectral broadening $\hbar\Gamma$ on the spectral resolution.

The absorption spectrum with no light-matter coupling is shown in Fig.\,\ref{fig:1}(c) and features a single dominant absorption peak corresponding to an electronic transition to $\ket{{\rm e}_1, 0}$ from $\ket{{\rm g}, 0}$, agreeing well with computational predictions in Ref.~\cite{flick2019lmrnqe} and with experiments presented in Ref.~\cite{benzeneAbsExp}. We verify that the excited eigenstate $\ket{{\rm e}_1, 0}$ is separated in energy at least 3 eV from the nearest eigenstates with substantial transition dipole moment in the $x$-direction. For the range of coupling rates considered in this paper, $\ket{{\rm g}, 0}$ and $\ket{{\rm e}_1, 0}$ thus effectively form a two-level electronic system for the considered range of parameters.


Next we place the molecule in a cavity whose central mode energy $\hbar\omega_c$ is in resonance with this transition ($\omega_{\rm c}=\omega_1^{\rm el}$). In the single excitation subspace, this system thus corresponds closely to the Jaynes-Cummings model, which describes the interaction of a two-level system with a single resonant bosonic mode. In this scenario schematically shown in Fig.\ref{fig:1}(a)-(b), the two-level system can be represented by a ground state $|{\rm g, 0}\rangle$ and an excited electronic state $|{\rm e}_1, 0\rangle$ of a molecule whose transition dipole moment $\bm{d}_1$ allows for efficient coupling with light and whose frequency is well separated from other electronic excitations. The bosonic mode is then represented by a single electromagnetic mode of an optical cavity (of quantized electric-field amplitude $\bm{E}_{\rm c}$ and experiences losses with decay rate $\kappa$) that interacts with the electronic states of the molecule via the Jaynes-Cummings coupling rate $g=-\frac{1}{\hbar}\bm{d}_1\cdot \bm{E}_{\rm c}$. If the coupling rate is small compared to the loss rate of the cavity, when $g<\kappa$, the system is in the weak-coupling regime in which the cavity enhances the decay rate of the electronic excited state, i.e. the Purcell effect. This enhancement manifests as broadening of the electronic absorption spectrum of the molecule as illustrated in Fig.\,\ref{fig:1}(a). In contrast, when $g>\kappa$, the system is in the strong-coupling regime in which the electronic excited state and the cavity photon hybridize and form new hybrid light-matter polariton states denoted $|U\rangle$, ``upper", and $|L\rangle$, ``lower", polariton in Fig.\,\ref{fig:1}(b). The polariton states are manifested in the molecular absorption spectrum as a doublet of peaks whose splitting is $\propto g$.

By tuning the cavity strength $\boldsymbol\lambda_{\rm c}$ (which we assume to be polarized in the $x$-direction) and loss $\hbar\kappa$, and therefore changing its spectral profile as shown in Fig.\,\ref{fig:cavityModes}(a), we are able to control the regime of light-matter coupling in accordance with the predictions of the Jaynes-Cummings model. We first calculate a series of absorption spectra of the molecule coupled with the cavity field:
In the calculations shown in Fig.\,\ref{fig:1}(d), $N=5,000$ cavity modes whose cavity strength is distributed according to Eq.\,\eqref{eq:lorentz} are used to represent the single leaky cavity mode coupled with a single dominant electronic excited state among the $M$ excited states of benzene, and the absorption spectrum is calculated with spectral broadening $\hbar\Gamma=0.001$ eV. 
 
We first assume that the cavity is characterised by a constant cavity loss $\hbar\kappa=0.001$\,eV and vary the value of cavity strength $\lambda_{{\rm c}}$. The calculated absorption spectra are shown in Fig.\,\ref{fig:1}(d)(i)-(iii) for $\lambda_{{\rm c}}=(0.001, 0.002,0.008)$\,eV$^{\frac{1}{2}}$/nm or $g/\kappa=(0.18,0.36,1.43)$, respectively. The largest $\hbar g=0.23$~meV $\ll 0.1\hbar\omega_1^{\rm el}$, so the coupling strength remains far below the ultrastrong regime~\cite{frisk2019ultrastrongreview, forndiaz2019uscRMP}. As we increase the coupling strength we observe the transition from the weak-coupling regime in Fig.\,\ref{fig:1}(d)(i) where the molecular absorption spectrum features a single broadened Lorentzian peak to the strong-coupling regime in Fig.\,\ref{fig:1}(d)(iii) featuring the polaritonic peak doublet. 
 
As the regime of light-matter coupling is a function of the ratio $g/\kappa$, we show in Fig.\,\ref{fig:1}(d)(iii)-(v) that the strong light-matter coupling regime can transition back into the weak-coupling regime when cavity losses are increased for a given cavity strength $\lambda_{{\rm c}}=0.008$ eV$^{\frac{1}{2}}$/nm. Concretely, we increase the cavity loss rate from $\hbar\kappa=0.001$\,eV in (iii) to $\hbar\kappa=0.004$\,eV in (iv) to $\hbar\kappa=0.008$\,eV in (v). We observe that with increasing loss rate the doublet merges back into a single broadened Lorentzian peak as we reenter the weak-coupling regime. The peak in Fig.\,\ref{fig:1}(d)(v) is broader than the peak in Fig.\,\ref{fig:1}(d)(i) as the Purcell decay $\propto g^2/\kappa$ is larger in the former case.   
 
To develop further physical intuition on how the $N$ photonic states hybridize with the $M$ electronic states to form $M+N$ polaritonic states, we calculate the weight $W_{il}^{\rm el}=|C_{il}^{\rm el}|^2$ ($W_{kl}^{\rm ph}=|C_{kl}^{\rm ph}|^2$) of an original, unmixed electronic (photonic) state $|{\rm e}_i,0\rangle$ ($|{\rm g},1_k\rangle$) in a resulting $l$th polaritonic state $| v_l\rangle$. We plot the total weight from the electronic states $w_l^{\rm el}=\sum_{i=1}^{M} W_{il}^{\rm el}$ and the total weight from the photonic states $w_l^{\rm ph}= \sum_{k=1}^{N} W_{kl}^{\rm ph}=1-w_l^{\rm el}$ as the overlaid color of the curve in Fig. \ref{fig:1}(d). Notably, we observe concentrations of electronic character near the absorption peaks, even when separated in energy as in Fig. \ref{fig:1}(d)(iii), which proves the hybrid polaritonic character of the states composing the peaks.

Finally we remark that in Appendix\,\ref{app:fano}, we show that results obtained from QEDFT in this parameter regime can be reproduced well using the cQED model, as expected in a model system.

\subsection{Many electronic excitations of toluene} \label{ssec:manylevel}
\begin{figure}[!htbp] 
\centering
\includegraphics[width=1.0\linewidth]{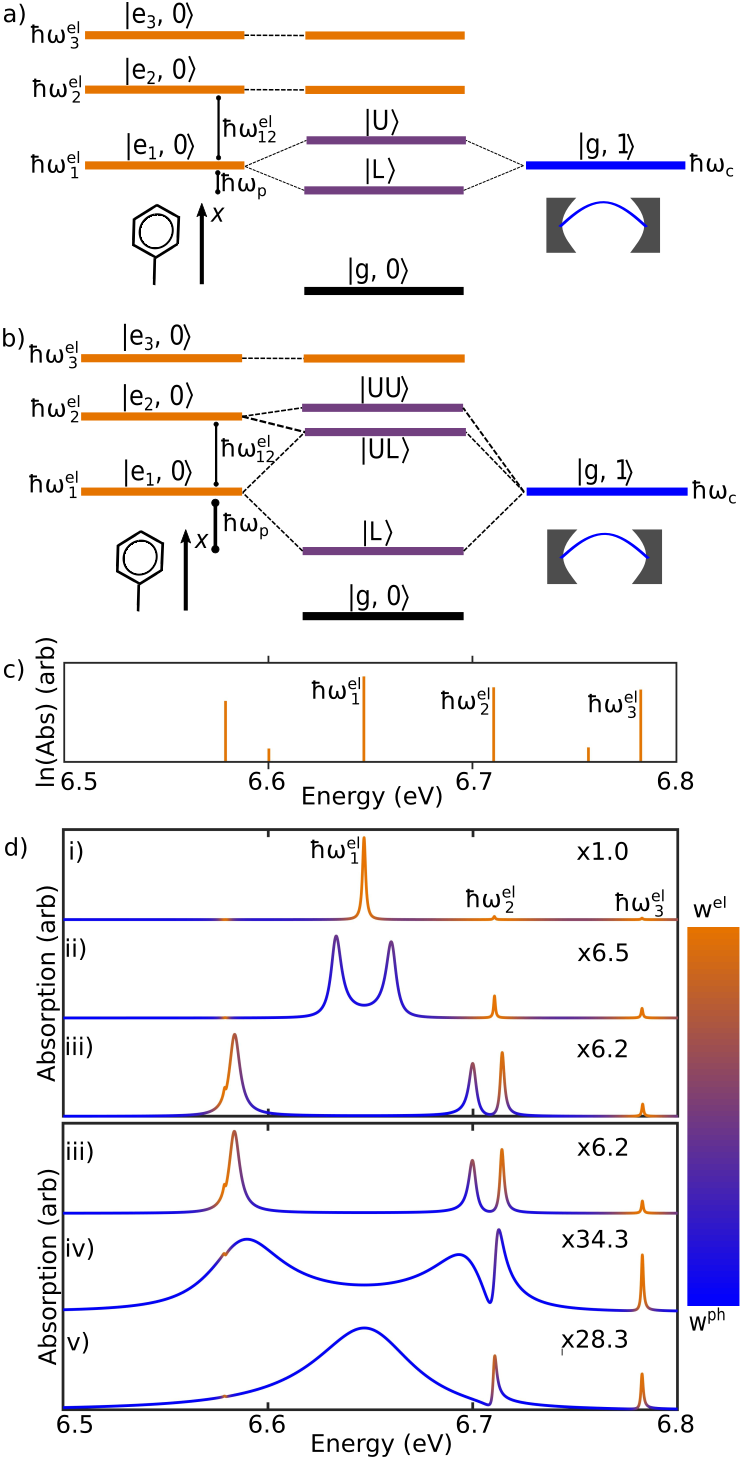}
\caption{ The ground state $\ket{{\rm g}, 0}$ and three excited electronic states--$\ket{{\rm e}_1, 0}$, $\ket{{\rm e}_2, 0}$, and $\ket{{\rm e}_3, 0}$, in order of increasing energy--of toluene comprise a many-level electronic system. \textbf{(a)} For coupling between $\ket{{\rm e}_1, 0}$ and a resonantly tuned, low loss optical cavity $\ket{{\rm g},1}$ where energy splitting $\omega_p \ll \omega_{12}^{\rm el}$ ($\omega_{12}^{\rm el}=\omega_2^{\rm el}-\omega_1^{\rm el}$), we observe splitting into upper and lower polaritons with little interaction with $\ket{{\rm e}_2, 0}$ and $\ket{{\rm e}_3, 0}$. \textbf{(b)} When the coupling strength is tuned such that $\omega_p \approx \omega_{12}^{\rm el}$, the upper polariton can be tuned in resonance and mixed with $\ket{{\rm e}_2, 0}$. \textbf{(c)} Normalized \textit{x}-polarized optical absorption spectrum of toluene in the absence of light-matter coupling. \textbf{(d)} Normalized optical \textit{x}-polarized absorption spectrum in the presence of light-matter coupling for spectral broadening $\hbar\Gamma$ = 0.001 eV. The coupling strength $\lambda_{\rm c}$ is set to (i) 0.01, (ii) 0.10, and (iii)-(v) 0.43 eV$^\frac{1}{2}$/nm. The loss $\hbar\kappa$ is set to (i)-(iii) 0.01, (iv) 0.10, and (v) 0.32 eV. The color of the curves corresponds to the relative photonic and electronic character of the polaritonic eigenstate. To retrieve the unnormalized absorption spectra, multiply the absorption intensities by the scaling factor on the right side of the plot.}
\label{fig:2}
\end{figure}

First principles-based techniques enable quantitative predictions \textit{a priori} of complex systems. Here, we systematically describe the effect of weak-to-strong light-matter coupling to an effective many-level electronic system, toluene. Although we again include all $M$ electronic excited states from the \textit{ab initio} calculation, here we discuss three $x$-polarized optically active excited states clustered in energy: $\ket{{\rm e}_1, 0}$ with energy $\hbar\omega_1^{\rm el}=6.64$ eV and $x$-polarized transition dipole moment $d_{1,x}=0.76$ e\AA, $\ket{{\rm e}_2, 0}$ with $\hbar\omega_2^{\rm el}=6.71$ eV and $d_{2,x}=0.11$ e\AA, and $\ket{{\rm e}_3, 0}$ with $\hbar\omega_3^{\rm el}=6.78$ eV and $d_{3,x}=0.08$ e\AA. The $x$-polarized absorption spectrum of toluene in free space is plotted in Fig. \ref{fig:2}(c) where we observe absorption peaks that scale as $\propto d_{i,x}^2$ corresponding to $\ket{{\rm e}_1, 0}$, $\ket{{\rm e}_2, 0}$, and $\ket{{\rm e}_3, 0}$. There are additional eigenstates that are further off-resonant or have smaller transition dipole moments. We neglect them in our discussion in the main text because they do not interact strongly with the cavity. Nonetheless, we still include their effects in our calculation. As we discuss in Appendix \ref{app:resolution}, for small enough spectral broadening $\hbar\Gamma$, these excitations appear in the spectra as absorption dips. 

When the molecule is placed in a 1D lossy cavity with central mode frequency $\omega_{\rm c}=\omega_1^{\rm el}$ and $M=36,000$ photon modes, we observe the transition from the weak-to-strong coupling regime in the electronic absorption spectra in Fig. \ref{fig:2}(d) with spectral broadening $\hbar\Gamma=0.001$ eV upon varying cavity strength $\lambda_{\rm c}$ and loss $\hbar\kappa$ as in Fig. \ref{fig:cavityModes}(b). At suitable cavity strength $\lambda_{\rm c}$ and loss $\hbar\kappa$, we also observe Fano resonances resulting from photon-mediated interactions between electronic states. 

We first assume that the cavity is characterized by cavity strength $\lambda_{\rm c}=0.01$eV$^\frac{1}{2}$/nm and loss $\hbar\kappa=0.01$ eV, or $g/\kappa=0.14$. As we increase the coupling  strength,  we  observe  the  transition  from  the weak-coupling  regime  in Fig. \ref{fig:2}(d)(i) where  the  molecular absorption spectrum features a broadened Lorentzian peak at $\hbar\omega_1^{\rm el}$ to  the  strong-coupling  regime in Fig. \ref{fig:2}(d)(iii) where $\lambda_{\rm c}=0.10$~eV$^\frac{1}{2}$/nm or $g/\kappa=1.4$ featuring the polaritonic peak doublet. This doublet corresponds to the lower $|{\rm L}\rangle$ and  upper $|{\rm U}\rangle$ polaritons in Fig. \ref{fig:2}(a). 

In a many-level electronic system, electronic states may interact through the photon modes. We demonstrate the effects of this interaction by increasing the cavity strength $\lambda_{\rm c}$ to 0.43 eV$^\frac{1}{2}$/nm and maintaining loss $\hbar\kappa= 0.01$ eV ($g/\kappa=6.0$) in Fig. \ref{fig:2}(d)(iii) from 0.10 eV$^\frac{1}{2}$/nm in Fig. \ref{fig:2}(d)(ii). At this cavity strength, $\omega_p\approx\omega_{12}^{\rm el}=\omega_2^{\rm el}-\omega_1^{\rm el}$, so the upper polariton becomes nearly degenerate $\ket{{\rm e}_2, 0}$ with energy $\hbar\omega_2^{\rm el}$, resulting in the upper polariton mixing with $\ket{{\rm e}_2, 0}$ to form an upper-lower polariton $|{\rm UL}\rangle$ and an upper-upper polariton $|{\rm UU}\rangle$. These conditions are schematically illustrated in Fig. \ref{fig:2}(b).

Increasing loss $\hbar\kappa$ for the toluene-lossy cavity system such that $\ket{{\rm e}_1, 0}$ and $\ket{{\rm e}_2, 0}$ interact through the photon modes results in Fano resonances in the absorption spectrum. To capture this effect, from the strong-coupling regime in Fig. \ref{fig:2}(d)(iii) where the cavity strength $\lambda_{\rm c}=0.43$ eV$^\frac{1}{2}$/nm and loss $\hbar\kappa=0.01$ eV, we increase loss $\hbar\kappa$ to 0.08 eV and maintain $\lambda_{\rm c}=0.43$ eV$^\frac{1}{2}$/nm, or $g/\kappa=0.74$, in Fig. \ref{fig:2}(d)(iv). Further increasing loss $\hbar\kappa$ enables resonant electronic eigenstates to interact with other electronic eigenstates even further in energy: in Fig. \ref{fig:2}(d)(v) where $\lambda_{\rm c}=0.43$ eV$^\frac{1}{2}$/nm and loss $\hbar\kappa$ is increased to 0.32 eV, the peak at $\hbar\omega_3^{\rm el}$ broadens. In addition, the individual polaritonic peaks merge into a single unresolved one, a characteristic of the weak-coupling regime. 

In Appendix\,\ref{app:fano} we discuss how the results obtained from QEDFT can be interpreted using the cQED model and compare the absorption spectra obtained from both models. We show that, up to small discrepancies arising from the rotating-wave approximation applied in cQED and the mean field approximation applied in QEDFT, as well as the exclusion of the $\bm{R}^2$ term arising from expansion of Eq. \eqref{eq:hamiltonian} in the cQED model, the two methods are in good quantitative agreement.

\begin{figure}
\centering
\includegraphics[width=1.0\linewidth]{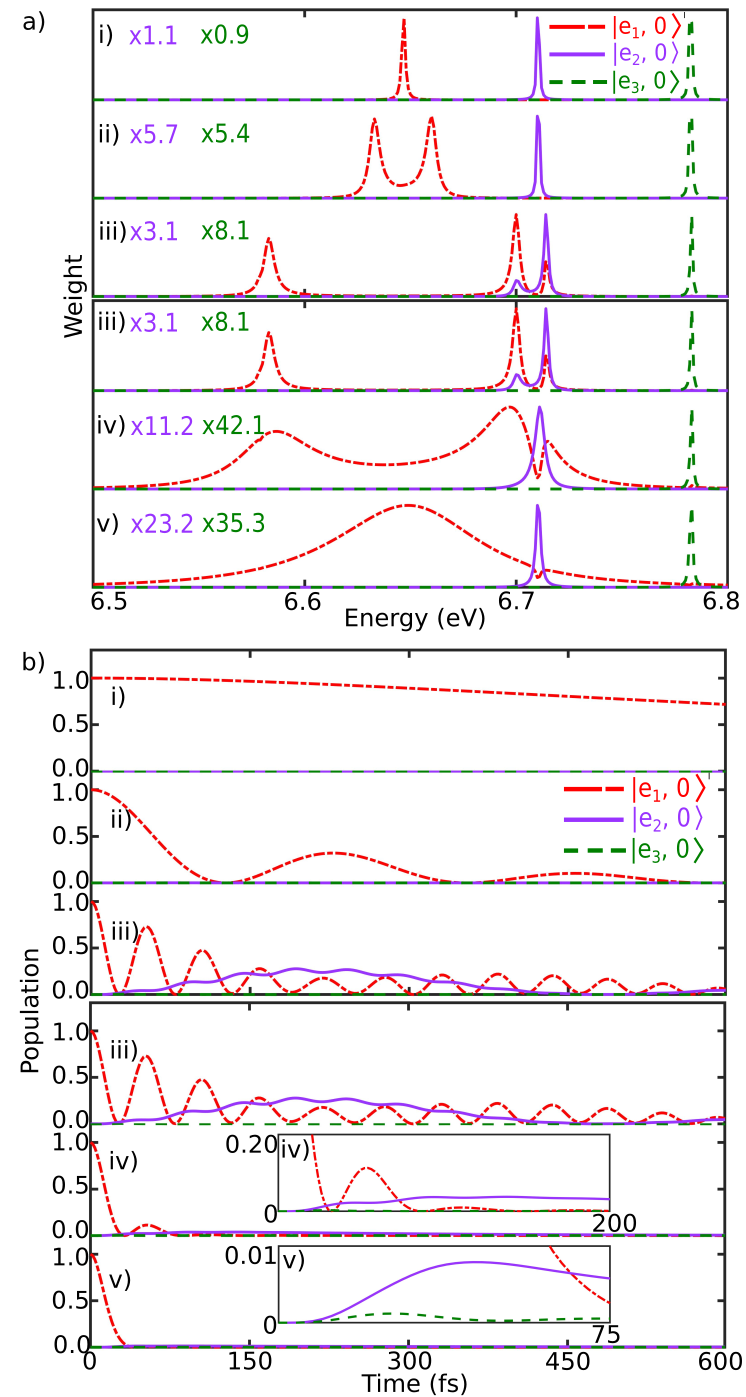}
\caption{
\textbf{(a)} Normalized weights of original, unmixed electronic states in the polaritonic states  calculated with QEDFT framework. All electronic states are decoupled from each other with low light-matter coupling in (i). The upper polariton resulting from mixing between $\ket{{\rm e}_1, 0}$ and $\ket{{\rm g}, 1}$ mixes with $\ket{{\rm e}_2, 0}$ in (iii). $\lambda$ and $\kappa$ for (i)-(v) in Fig. \ref{fig:3}(a) equal those in (i)-(v) in Fig. \ref{fig:2}(d). The unnormalized weights are recovered by scaling the plotted curves by the listed scale factors. 
\textbf{(b)} Population transfer in the time domain via a cQED model with inputs from first principles calculations of the excited electronic eigenstates and photonic spectral profile. Toluene is initially prepared in $\ket{{\rm e}_1, 0}$.  Rabi oscillations in (ii) are a clear indication of strong coupling between $\ket{{\rm e}_1, 0}$ and $\ket{{\rm g},1}$. Population transfer between $\ket{{\rm e}_1, 0}$ and $\ket{{\rm e}_2, 0}$ occurs when $\omega_p \approx \omega_{12}^{\rm el}$, as in (iii) and schematically illustrated in Fig. \ref{fig:2}(b). Increasing $\kappa$ in (iv)-(v) decreases excited electronic state lifetimes and enables slight population transfer to $\ket{{\rm e}_3, 0}$, as observed in (v). $\lambda$ and $\kappa$ for (i)-(v) in Fig. \ref{fig:3}(b) equal those in (i)-(v) in Fig. \ref{fig:2}(d).}
\label{fig:3}
\end{figure}

To elucidate the physical implications of the absorption spectra, we plot the weights $W_{il}^{\rm el}$
from QEDFT calculations in Fig. \ref{fig:3}(a)(i)-(v). The conditions correspond to the same cavity conditions as in Fig. \ref{fig:2}(d)(i)-(v), respectively, except that the input cavity mode frequency range is restricted from 6.35 eV to 6.95 eV in Fig. \ref{fig:3}(a)-(b) for computational tractability as opposed to 4.85 eV to 8.45 eV in Fig. \ref{fig:2}(d). Absorption spectra are nearly identical for both energy ranges. Since the absorption spectra calculated from first principles in Fig. \ref{fig:2}(d) agree with those of the cQED model in Fig. \ref{fig:fano} parameterized with first principles calculations of the electronic structure and cavity mode profiles, we can use the cQED model to calculate the explicit time dependence of population transfer in Fig. \ref{fig:3}(b) for a toluene-lossy cavity system initially prepared in $\ket{{\rm e}_1, 0}$.

In the weak-coupling regime in Fig. \ref{fig:3}(a)(i), we see that all electronic character from $\ket{{\rm e}_1, 0}$, $\ket{{\rm e}_2, 0}$, and $\ket{{\rm e}_3, 0}$ is localized in one peak each. To demonstrate weak-coupling behavior in the time domain, the time decay of the toluene-lossy cavity system initialized in $\ket{{\rm e}_1, 0}$ is plotted in Fig. \ref{fig:3}(b)(i). As expected for an electronic system weakly coupled to photon modes, we observe exponential decay of the population of $\ket{{\rm e}_1, 0}$.

In the strong-coupling regime where the upper $|{\rm U}\rangle$ and lower $|{\rm L}\rangle$ polaritons are well-resolved in Fig. \ref{fig:2}(d)(ii), in Fig. \ref{fig:3}(a)(ii) the electronic character of $\ket{{\rm e}_1, 0}$ is concentrated within the two peaks corresponding to $|{\rm U}\rangle$ and $|{\rm L}\rangle$. We calculate the time decay of the toluene-lossy cavity system initialized in $\ket{{\rm e}_1, 0}$ in Fig. \ref{fig:3}(b)(ii) for the same cavity parameters as in Fig. \ref{fig:2}(d)(ii) and observe a signature of strong-coupling behavior: vacuum Rabi oscillations. In this regime, the difference in energy $\hbar\omega_{\rm p}$ between the polaritons and $\ket{{\rm e}_1, 0}$ is much lower than the difference in energy $\hbar\omega_{12}^{\rm el}$ between $\ket{{\rm e}_1, 0}$ and $\ket{{\rm e}_2, 0}$, so there is negligible interaction between $\ket{{\rm e}_1, 0}$ and $\ket{{\rm e}_2, 0}$. This lack of interaction is apparent in both Fig. \ref{fig:3}(a)(ii) and Fig. \ref{fig:3}(b)(ii). In Fig. \ref{fig:3}(a)(ii), the two polaritonic peaks have no contribution from $\ket{{\rm e}_2, 0}$, and in Fig. \ref{fig:3}(b)(ii), we observe no population transfer to $\ket{{\rm e}_2, 0}$.

In Fig. \ref{fig:2}(d)(iii), $\lambda_{\rm c}$ is further increased such that electronic states $\ket{{\rm e}_1, 0}$ and $\ket{{\rm e}_2, 0}$ interact. In Fig. \ref{fig:3}(a)(iii), the peaks corresponding to $|{\rm UU}\rangle$ and $|{\rm UL}\rangle$ have contributions from both $\ket{{\rm e}_1, 0}$ and $\ket{{\rm e}_2, 0}$. The consequence in the time domain is population transfer between $\ket{{\rm e}_1,0}$ and $\ket{{\rm e}_2,0}$, as well as Rabi oscillations with the photon modes, suggesting that this toluene-lossy cavity system experiences both strong light-matter and strong matter-matter coupling via the cavity modes. As loss $\hbar\kappa$ increases to 0.10 eV in Fig. \ref{fig:2}(d)(iv), although population still transfers between $\ket{{\rm e}_1, 0}$ and $\ket{{\rm e}_2, 0}$, the excitation lifetime commensurately decreases compared to when $\hbar\kappa=0.01$ eV. Finally, as $\kappa$ is increased further to 0.32 eV in Fig. \ref{fig:2}(d)(v) where $\ket{{\rm e}_3, 0}$ interacts non-negligibly with the cavity modes, we observe slight population transfer to $\ket{{\rm e}_3, 0}$ in Fig. \ref{fig:3}(b)(v).

Overall, we demonstrate that tuning the cavity strength $\lambda_{\rm c}$ and loss $\hbar\kappa$ can generate polaritonic states with contributions from several excited electronic states and can control population transfer to higher-lying states in energy and excitation lifetimes. 


\section{Conclusions} \label{discussion}
To summarize, we study \textit{ab initio} correlated optical interactions in matter ranging from the weak-coupling to strong-coupling regime. As an example, we calculate excited-state dynamics and spectral responses of benzene and toluene as effective two-level and many-level electronic systems, respectively, under variable light-matter coupling controlled by the cavity strength $\lambda_{\rm c}$ and loss $\hbar\kappa$. By tuning the cavity parameters, we notice transitions between the weak-coupling and the strong-coupling regimes where polaritonic states can be resolved. In the many-level electronic system, we observe Fano resonances in the electronic absorption spectrum resulting from interactions between electronic excited states mediated by the cavity. These interactions enable cavity-mediated population transfer between electronic excited states where the lifetimes and degree of population transfer are controlled by the cavity parameters. We reproduce the first principles results using a cQED model parametrized with the QEDFT data and generally note excellent agreement.

This generalized QEDFT formalism is especially useful for predictions involving molecules with complex excited state profiles interacting with arbitrary electromagnetic environments, which can be parameterized by, for instance, cavity strength, loss, photon mode density, and dimensionality of cavity modes of photonic crystals and plasmonic cavities. Looking forward, in electronic structure theory, we anticipate that the development of improved TDDFT methods will improve prediction accuracy for higher-lying excited states. In cQED, further understanding the effects of the correct inclusion of the $\bm{R}^2$ term in the QEDFT formalism versus a conventional cQED model invoked in the present study will enable the extension of this method to the ultrastrong coupling regime.

Looking forward, this work is an important predictive technique for experiments involving the weak- and strong-coupling light-matter regimes, including modifications of the excited state potential energy surfaces in the field of polaritonic chemistry. This extension to QEDFT moves toward closing the loop between first principles calculations in electronic structure theory and parametric models of the quantum optics community.

\section*{Acknowledgements}
This work was supported by the DOE `Photonics at Thermodynamic Limits’ Energy Frontier Research Center under grant DE-SC0019140. DW is an NSF Graduate Research Fellow. The Flatiron Institute is a division of the Simons Foundation. PN is a Moore Inventor Fellow.

\appendix

\section{Fano-type cQED model} \label{app:fano}

\begin{figure*}
\centering
\includegraphics[width=1.0\linewidth]{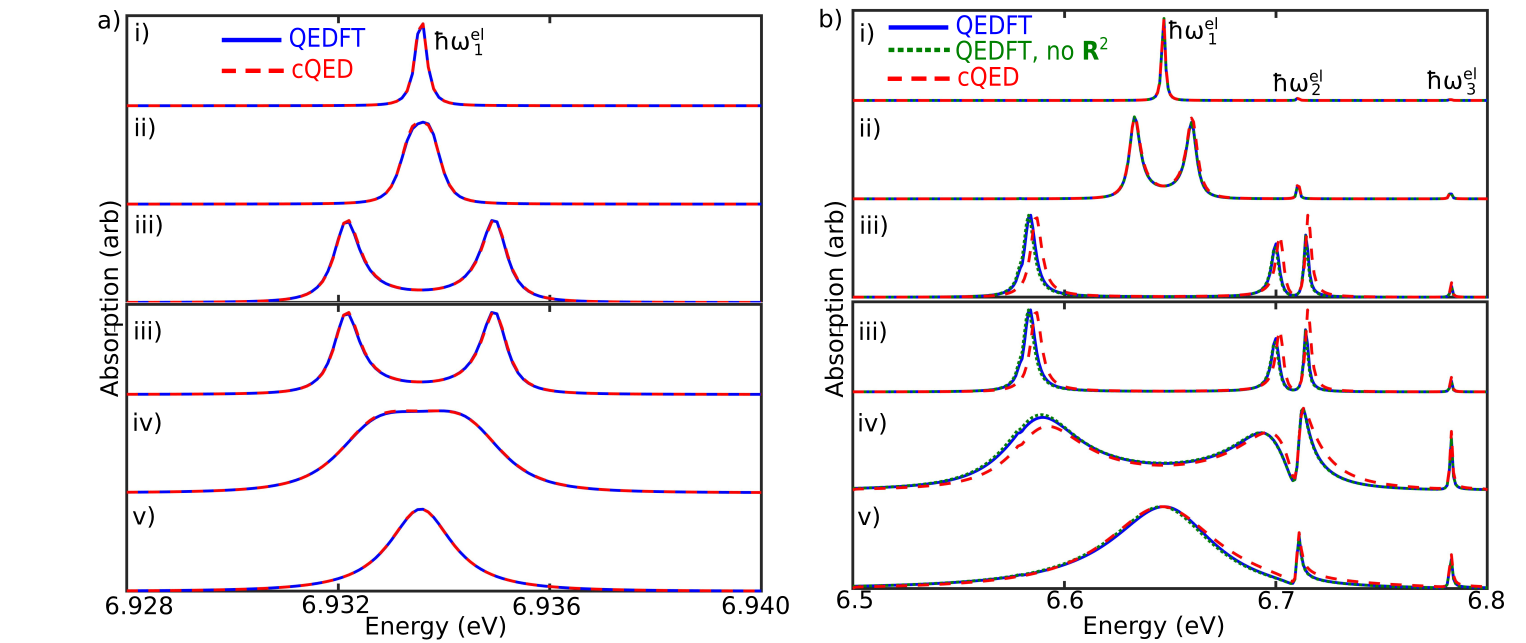}
\caption{Comparison of normalized $x$-polarized absorption spectra for \textbf{(a)} benzene and \textbf{(b)} toluene between QEDFT and a cQED model parameterized with the eigenenergies and transition dipole moments of the electronic excited states and the same respective cavity mode profiles as the first principles, self-consistent QEDFT calculations. QEDFT curves of Fig. \ref{fig:fano}(a) and Fig. \ref{fig:fano}(b) are re-plotted from Fig. \ref{fig:1}(d) and Fig. \ref{fig:2}(d), respectively.}
\label{fig:fano}
\end{figure*}

The Fano-type cQED model describes the linear-response of a discrete many-level electronic system coupled to a continuous photonic reservoir, permitting analysis in both the time and frequency domains. The relevant Hamiltonian $H_{\rm cQED}$, in the rotating wave approximation, is given by
\begin{multline} \label{app:eq1}
H_{\rm cQED}=\sum_{i=1}^{M}\hbar\omega^{\rm el}_{i}\sigma_i^\dagger\sigma_i + \sum_{k=1}^{N}\hbar \omega_k a^\dagger_k a_k
\\+ \hbar\sum_{i,k=1}^{M,N} (g_{i,k} \sigma_i^\dagger a_k+{\rm H.c.}),
\end{multline}
where $\sigma_i^\dagger$ ($\sigma_i$), $a_k^\dagger$ ($a_k$) are creation (annihilation) operators for the $i$th of $M$ excited electronic states and $k$th of $N$ photon modes, respectively, H.c. is the Hermitian conjugate, $\hbar\omega^{\rm el}_{i}$ and $\hbar\omega_{k}$ are the mode energies, and $g_{i,k}$ is the coupling defined as in Eq. \eqref{eq:g}.



The state of the system can be described in the single excitation subspace as
\begin{align} \label{app:eqFano1}
    |\psi\rangle=& c_0|{\rm g},\{0_k\}\rangle+\sum_{i=1}^{M}c_{i}^{\rm el}|{{\rm e}_i}, \{0_k\}\rangle+\sum_{k=1}^{N}c_{k}^{\rm ph}|{\rm g}, \{1_k\}\rangle,
\end{align}
where the coefficients $c_0$, $c_i^{\rm el}$, $c_k^{\rm ph}$  are time dependent. We plug this \textit{ansatz} into the Schr\"{o}dinger equation to obtain the following system of linear differential equations:
\begin{align} \label{app:eqFano2}
    \dot c_{i}^{\rm el} &=-{\rm i}\omega^{\rm el}_i c_{i}^{\rm el}-{\rm i}\sum_{k=1}^{N} g_{i,k} c_k^{\rm ph},\\
    \dot c_k^{\rm ph} &=-{\rm i}\sum_{i=1}^{M}g^\ast_{i,k} c_{i}^{\rm el}-{\rm i}\omega_k c_k .
\end{align}

Note that the coefficient $c_0$ drops from the system of equations as the Hamiltonian conserves the number of excitations due to the rotating wave approximation. This system of linear differential equations can be solved to obtain the eigenvalues, $\hbar\omega_l$ (i.e. energies), and eigenvectors, $|v_l\rangle$ (i.e. polaritonic states), of the coupled system. 

With these parameters, we can calculate the electronic absorption spectrum and weight of the original, unmixed electronic and photonic states as discussed in Section \ref{ssec:twolevel}. Normalized $x$-polarized absorption spectra for benzene and toluene with the same cavity conditions in Fig. \ref{fig:cavityModes}(a) and Fig. \ref{fig:cavityModes}(b) are plotted in Fig. \ref{fig:fano}(a) and Fig. \ref{fig:fano}(b), respectively. We note that the QEDFT and cQED agree quantitatively well for both benzene and toluene. There are more substantial differences between the spectra, specifically a blue shift of the cQED spectra, for the toluene study, which used higher coupling strengths than the benzene study. This trend agrees with a comparison between QEDFT and the Rabi model in Ref.~\cite{flick2019lmrnqe}. To understand the effect of the lack of the $\bm{R}^2$ term in the cQED model that arises from expanding the last term in Eq. \eqref{eq:hamiltonian}, in Fig. \ref{fig:fano}(b), we also plot the absorption spectra calculated with the present QEDFT formalism without the $\bm{R}^2$ term. The effect of the $\bm{R}^2$ is small for the coupling strengths here, so the differences in spectra can be more directly attributed to differences in the rotating wave approximation made in the cQED model and the mean-field approximation made in QEDFT. Finally, Eq. \eqref{app:eqFano1} and Eq. \eqref{app:eqFano2} can also be explicitly propagated in time given a set of initial conditions to plot population transfer, as in Fig. \ref{fig:3}(b).

\section{Transformation of photon mode density} \label{app:transform}
Here we demonstrate how to transform the frequency spacing of a photon spectral profile from an exact calculation of the electronmagnetic environment of a cavity to an arbitrary mode frequency spacing. Such an equivalency has clear numerical advantages whereby frequencies of interest can be more densely sampled at the cost of lowered frequency density at less relevant ranges for the same computational cost.

For specificity, we study the Hamiltonian $H_{\rm cQED, cont}$ in the cQED model, Eq. \eqref{app:eq1}, expressed in terms of a continuously, constantly spaced frequency variable $\omega$ for photons and discrete frequencies $\omega_{i}^{\rm el}$ for electronic excited states:

\begin{multline} \label{app:eq3}
H_{\rm cQED,cont}=\sum_{i=1}^{M}\hbar\omega^{\rm el}_{i}\sigma_i^\dagger\sigma_i + \hbar \int {\rm d} \omega\, \omega a^\dagger(\omega) a(\omega) \\
+ \hbar\int {\rm d} \omega \sum_{i=1}^{M} (g_{i}(\omega) \sigma_i^\dagger a(\omega)+{\rm H.c.}).
\end{multline}

We desire to change the integration variable $\omega$ to a new one $\Omega$ such that $\omega(\Omega)$ has a more favorable spacing once the Hamiltonian is necessarily discretized for numerical calculations. Assuming a general functional dependence connecting the two frequency variables, note that the transformation of variables leads to modification of all integrals:
\begin{align}
    \int \ldots {\rm d}\omega \to \int \ldots \underbrace{\frac{{\rm d}\omega}{{\rm d}\Omega}}_{\equiv D(\omega\to\Omega)}{\rm d}\Omega,
\end{align}
where we have defined the density of states $D(\Omega)$. This transformation is accompanied by re-definition of creation and annihilation operators of the continuum:
\begin{align}
    \tilde a(\omega(\Omega))=\sqrt{D(\Omega)} a(\omega),
\end{align}
which follows from the requirement that the commutation relation $[\tilde a(\Omega), \tilde a ^\dagger(\Omega ')]=\delta(\Omega-\Omega')$ holds.

We can rewrite $H_{\rm cQED, cont}$ as
\begin{multline}
    H_{\rm cQED, cont}= \\
    \sum_{i=1}^{M}\hbar\omega_{i}^{\rm el}\sigma_i^\dagger\sigma_i + \hbar\int {\rm d}\Omega\, \omega(\Omega) \tilde a^\dagger (\Omega)\tilde a (\Omega) \\
    +\hbar \int {\rm d}\Omega\,\sum_{i=1}^{M} \left[\underbrace{\sqrt{D(\Omega)}G_i[\omega(\Omega)]}_{\equiv \tilde G_i(\Omega)}\sigma_i^\dagger \tilde  a(\Omega)+ {\rm H.c.}\right], \label{app:eq2}
\end{multline}
where we have defined a new coupling constant $\tilde G_i(\Omega)$. Eq. \eqref{app:eq2} can be discretized by assuming a constant spacing in variable $\Omega$: 
\begin{align}
    \int {\rm d}\Omega \to \sum_{k=1}^{N} \Delta\Omega_k ,
\end{align}
where $\Delta\Omega=\Omega_{i+1}-\Omega_{i}$ is the step of the discretization. The annihilation and creation operators of the continuum must be transformed back to the operators of the discrete set of modes as:
\begin{align}
    \tilde A(\Omega_i)\equiv \tilde A_k \equiv \sqrt{\Delta\Omega}\tilde a(\Omega_k),
\end{align}
which follows from the requirement that $[\tilde A_i, \tilde A^\dagger_j]=\delta_{ij}$.
We write the discrete version of $H_{\rm cQED,cont}$, $\tilde H_{\rm cQED}$, with new mode frequency spacing:
\begin{multline}
        \tilde H_{\rm Fano}=\sum_{i=1}^{M}\hbar\omega^{\rm el}_{i}\sigma_i^\dagger\sigma_i +\sum_{k=1}^{N}\hbar\omega(\Omega_k)\,\tilde A^\dagger_k \tilde A_k \\
    +\hbar\sum_{i,k=1}^{M,N}\,  \left[ \sqrt{\Delta \Omega}\tilde G(\Omega_k)\sigma_i^\dagger \tilde A_k + {\rm H.c.}\right].
\end{multline}


\section{Computational details of TDDFT/QEDFT} \label{app:converge}

We use a locally modified version of the pseudopotential, real-space DFT code OCTOPUS~\cite{octopus1, octopus2,octopus3}. The calculation of electronic excited states via the Casida TDDFT method first requires a converged ground state electronic structure on optimized molecular geometries. Molecular geometries are optimized with norm-conserving, plane-wave Hamann, Schl{\"u}ter, Chiang, and Vanderbilt (HSCV) pseudopotentials~\cite{pseudodojo} and an local density approximation (LDA) functional~\cite{LDA}. The real-space simulation box is parameterized with a mesh spacing of 0.16~\AA~and consists of spheres with 6~\AA~radius around each atom.

We calculate the converged ground state electron densities with a standard LDA functional for benzene and a long-range adapted LDA functional for toluene~\cite{longrangeLDA}; the long-range LDA (LRLDA) functional requires the ionization potential from a $\Delta$SCF calculation. Ground state self-consistent field (SCF) energies converged within $\sim1$ meV/atom with 0.08~\AA~and 0.14~\AA~real-space mesh spacing for benzene and toluene, respectively, and a simulation box consisting of 6~\AA~radius around each atom.

The energies of the excited states of interest converged within $\sim1$ meV/atom with 500 extra states in the Casida method. We further confirm the reliability of the excited electronic eigenstates that we couple to the cavity modes by noting that all excited states studied lie below the ionization potentials determined with a $\Delta$SCF calculation.

\section{Spectral broadening} \label{app:resolution}
\begin{figure}[tbhp]
\centering
\includegraphics[width=1.0\linewidth]{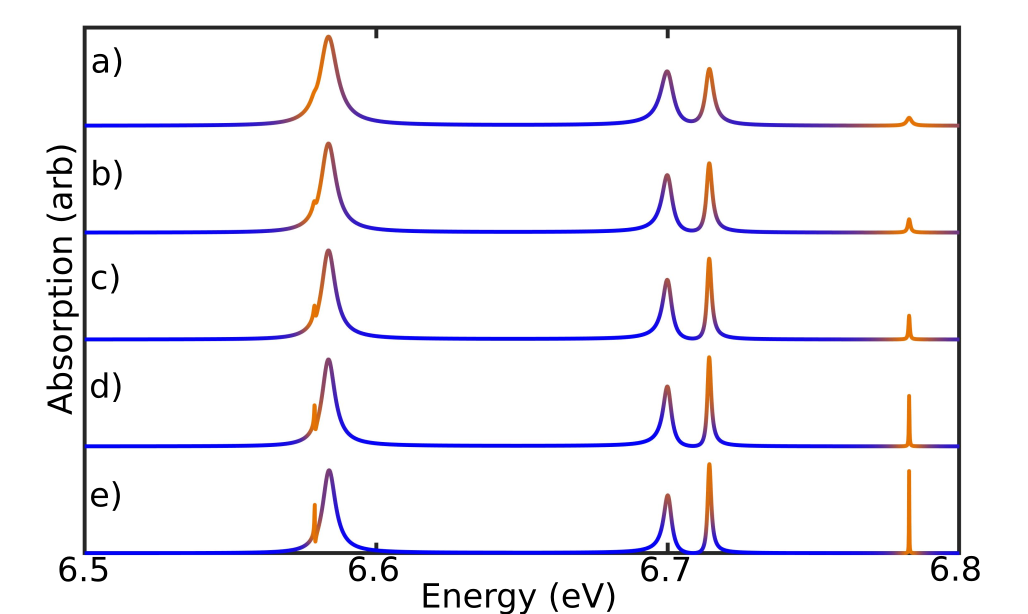}
\caption{Normalized $x$-polarized absorption spectra for toluene in a lossy cavity with $\lambda_{\rm c}=0.43$ eV$^{\frac{1}{2}}$/nm and $\hbar\kappa=0.01$ eV with spectral broadening $\hbar\Gamma=(0.005, 0.002, 0.001, 0.0005, 0.0002)$ eV for (a)-(e), respectively. Curve (c) corresponds to Fig. \ref{fig:2}(d)(iii).}
\label{fig:resolution}
\end{figure}

Molecules can have excitations densely spaced in energy that all interact through the cavity photons. However, because excitations with low transition dipole moments or that are sufficiently off-resonant from the line width of the cavity modes do not interact strongly with the cavity, the effects of their interactions are small. Nonetheless, it is possible to observe their effects by increasing measurement quality. 

For instance, for toluene in the energy range shown in Fig. \ref{fig:2}(c), there are several excitations beyond $\ket{{\rm e}_1, 0}$, $\ket{{\rm e}_2, 0}$, and $\ket{{\rm e}_3, 0}$ that do not have an appreciable effect on the absorption spectra when coupled to photons, as in Fig. \ref{fig:2}(d). The only observable impact is a slight dip on the side of the lower polariton in Fig. \ref{fig:2}(d)(iii)-(v), caused by the eigenstate at 6.58 eV with transition dipole moment $d_x=0.01$ e\AA. We study the impact of varying the spectral broadening $\hbar\Gamma$ on the absorption spectra for the conditions in Fig. \ref{fig:2}(c), as demonstrated in Fig. \ref{fig:resolution}. Physically, the spectral broadening can be lowered by decreasing the temperature. As the spectral broadening $\hbar\Gamma$ decreases from $\hbar\Gamma=(0.005, 0.002, 0.001, 0.0005, 0.0002)$ eV in Fig. \ref{fig:resolution}(a)-(e), respectively, we observe the appearance of an additional Fano resonance at the eigenenergy at $6.58$\,eV, a signature of the presence of an electronic eigenstate interacting with others through the photon modes. Given sufficient measurement resolution, we expect that further decreasing the spectral broadening $\hbar\Gamma$ would enable the observation of the remaining eigenstates in Fig. \ref{fig:2}(c).



\bibliography{biblio}
\end{document}